\documentclass[manuscript,screen]{acmart}

\usepackage{algorithm}
\usepackage{algorithmic}
\usepackage{amsbsy}
\usepackage{pifont}
\AtBeginDocument{%
  \providecommand\BibTeX{{%
    \normalfont B\kern-0.5em{\scshape i\kern-0.25em b}\kern-0.8em\TeX}}}

\setcopyright{acmcopyright}
\acmJournal{JETC}
\acmYear{2022} \acmVolume{1} \acmNumber{1} \acmArticle{1} \acmMonth{1} \acmPrice{15.00}\acmDOI{10.1145/3517810}

\newif\ifrev
	\revtrue
	\ifrev
	\newcommand{\yier}[1]{{\color{blue} [Yier: #1]}}
	\newcommand{\maxp}[1]{{\color{red} [Max: #1]}}
	\newcommand{\honggang}[1]{{\color{cyan} [Honggang: #1]}}
	\else
	\newcommand{\directions}[1]{}
	\newcommand{\yier}[1]{}
	\newcommand{\maxp}[1]{}
	\newcommand{\honggang}[1]{}
	\fi

\begin{document}

\title{A Review and Comparison of AI Enhanced Side Channel Analysis}
\author{Max Panoff}
\email{m.panoff@ufl.edu}
\orcid{0000-0003-2849-7197}
\affiliation{%
  \institution{University of Florida}
  \streetaddress{P.O. Box 1212}
  \city{Gainesville}
  \state{FL}
  \country{USA}
  \postcode{32611}
}

\author{Honggang Yu}
\email{honggang.yu@ufl.edu}
\orcid{0000-0002-9134-206X}
\affiliation{%
  \institution{University of Florida}
  \streetaddress{P.O. Box 1212}
  \city{Gainesville}
  \state{FL}
  \country{USA}
  \postcode{32611}
}

\author{Haoqi Shan}
\email{haoqi.shan@ufl.edu}
\orcid{0000-0003-1440-1828}
\affiliation{%
  \institution{University of Florida}
  \streetaddress{P.O. Box 1212}
  \city{Gainesville}
  \state{FL}
  \country{USA}
  \postcode{32611}
}

\author{Yier Jin}
\orcid{0000-0002-8791-0597}
\email{yier.jin@ece.ufl.edu}
\affiliation{%
  \institution{University of Florida}
  \streetaddress{P.O. Box 1212}
  \city{Gainesville}
  \state{FL}
  \country{USA}
  \postcode{32611}
}

\renewcommand{\shortauthors}{Panoff et al.}

\thanks{Corresponding Author: Yier Jin}

\begin{abstract}
Side Channel Analysis (SCA) presents a clear threat to privacy and security in modern computing systems. The vast majority of communications are secured through cryptographic algorithms. These algorithms are often provably-secure from a cryptographical perspective, but their implementation on real hardware introduces vulnerabilities. Adversaries can exploit these vulnerabilities to conduct SCA and recover confidential information, such as secret keys or internal states. The threat of SCA has greatly increased as machine learning, and in particular deep learning, enhanced attacks become more common. 
In this work, we will examine the latest state-of-the-art deep learning techniques for side channel analysis, the theory behind them, and how they are conducted. 
Our focus will be on profiling attacks using deep learning techniques, but we will also examine some new and emerging methodologies enhanced by deep learning techniques, such as non-profiled attacks, artificial trace generation, and others. 
Finally, different deep learning enhanced SCA schemes attempted against the ANSSI SCA Database (ASCAD) and their relative performance will be evaluated and compared. This will lead to new research directions to secure cryptographic implementations against the latest SCA attacks. 
\end{abstract}

\ccsdesc[500]{Hardware~Integrated circuits}
\ccsdesc[500]{Security and privacy~Cryptanalysis and other attacks}
\ccsdesc[500]{Computing methodologies~Artificial intelligence}

\keywords{Deep Learning, Side Channel Analysis, Cryptographic Hardware}

\maketitle

\section{Introduction}
\label{sec:intro}

Side Channel Analysis (SCA) is a family of techniques in which various physical aspects of a device, such as power consumption \cite{DPA}, ElectroMagnetic (EM) emanations \cite{agrawal2002side}, time to execute \cite{coppens2009practical}, and others are analyzed to break confidentiality or recover device operations. SCA can be used even against provably secure algorithms as the leakage comes from the hardware running the algorithm, rather than the algorithm itself. These cryptography algorithms are central to maintaining secure and private communications in our current world. As a result, SCA poses a significant and present threat that must be understood and addressed.
 
SCA attacks are increasingly targeting Internet of Things (IoT) devices and components, such as cryptographic hardware on System-on-a-Chips (SoCs), Field Programmable Gate Arrays (FPGAs), and Application Specific Integrated Circuits (ASICs), as well as on microprocessors. In particular, power and EM based SCA is growing in popularity and capability. In these techniques, adversaries collect power consumption or electromagnetic emanation measurements, known collectively as \textit{traces}, from devices performing operations. By analyzing these traces, the adversary gains some insight into the operations being performed, and then leverages this insight to recover confidential information.

At the same time, machine learning has become an increasingly popular solution to various problems, most notably image recognition \cite{lin2017focal, szegedy2017inception, wang2016cnn, deng2009imagenet} and natural language processing \cite{ruder2019transfer, otter2020survey, young2018recent, gardner2018allennlp}. Due to its strong performance in those domains, researchers have explored applying machine learning to many other new areas. Hardware security holds many analysis-heavy problems which machine learning is well suited to solve. As a result, researchers have created a variety of deep learning for side channel analysis methodologies. These deep learning based SCA methods can be highly effective, often demonstrating significant benefits over their traditional statistical counterparts. As such, we believe that an accounting of these techniques, their strengths and weakness, and a thorough analysis of what adversaries need to conduct them will inform future work in this area. 

Deep learning (DL) enhanced SCA can take a variety of forms but often are a variation of the Profiling Attack~\cite{maghrebi2016breaking}, a type of Template Attack \cite{chari2002template}.
As an example of this attack, let us assume that an adversary is targeting the key on a device running the Advanced Encryption Standard (AES) algorithm.
The attacker then needs access to an device identical to the victim device, through which they build a profile. The attacker has complete control over this \textit{profiling device}, including the ability to choose plaintext or ciphertext. Because of this, the attacker can use the plaintext or knowledge of the algorithm to create labels for machine learning. The attacker can collect side channel information from the profiling device, called \textit{traces} under predetermined inputs. The collected traces are labelled with Key-Text Pairs (KTPs). By training a DL network on these traces, a model that predicts label for unseen traces belonging to that same device can be created. At this point, the attacker needs only the ability to collect additional traces from the victim device along with knowledge of the plaintext used with each trace to recover the secret key.

Additionally, it has been proved possible to transfer DL SCA attacks between devices, something impossible with traditional approaches \cite{yucross, warriors, das2019x}. 
The transferred attack greatly expands an attack's capabilities, since far less data is needed to adapt a model to a new device rather than build one. In \cite{ASCAD} the authors find that models built from scratch need around 50000 profiling traces to recover the key within 1000 attacking traces. Even further, in \cite{yucross} the authors find that a model can be transferred to a new device, architecture, or side channel domain (e.g power, EM) with only $\sim$800 additional traces. As a result, implementations may be far more vulnerable to SCA than previously assumed.

As SCA attacks have been well explored, there has been significant work accomplished on defences and mitigation techniques against it. There are many different forms these defences can take, but they are often classified into two groups, Masking or Hiding. Masked implementations never handle data directly, creating an additional hurdle to attackers.
Hiding techniques on the other hand seek to limit and lower the amount of side channel leakage that occurs. DL SCA has already been shown to defeat a number of approaches belonging to each strategy \cite{ASCAD, gilmore2015neural}.

These new developments have led us to present this survey in order to explain some the most impactful Deep Learning (DL) Side Channel Analysis techniques. 
We will also discuss their weaknesses, and particularly their methodologies and the requirements to conduct these attacks. Through this, we hope that defenders will gain insight into how to better protect their devices while also providing useful information to those entering the field. 
Since many of these works use the ANSSI SCA Database (ASCAD), we will further compare the effectiveness and capabilities of existing DL SCA attacks. 
As they all use the same dataset, differences between the approaches will be highly informative. Through this analysis, we will also provide insights into the aspects of SCA that have not received as much attention, and how they can be and are being improved by Deep Learning.
Overall, through this survey on the newly proposed AI enhanced SCA attacks, we hope hardware security researchers would take the new threat into consideration when developing effective defense. 

Our paper will be organized as follows. In this Section, Section \ref{sec:intro}, we provide a brief introduction to our work. Section \ref{sec:background} contains a discussion of deep learning and side channel analysis to provide background to unfamiliar readers. We will discuss the latest DL SCA techniques in Section \ref{sec:prof} and in Section \ref{sec:Comparisons} we will compare these the results these {techniques report when applied on an open-source database}. A discussion of new and emerging approaches to DL SCA can be found in \ref{sec:emerging}. Finally, in Section \ref{sec:Conclusion} we present our final conclusions.

\section{Background}
\label{sec:background}

\subsection{Side Channel Analysis}

Side channels can take many forms, including power consumption \cite{DPA}, EM emanations  \cite{agrawal2002side}, time to execute \cite{coppens2009practical}, and others. Of these, the power and EM side channels are often the focus, because of the level of insight they provide, but also because they are heavily correlated. Both leakages are caused by the electrical switching of a design's internal gates. Changes in the state of these gates consumer power and release EM emanations from the device.   As the power and EM fluctuations depend on the state, different operations or data values have different impacts on these measurements. Identifying the proper time ranges to collect data can prove difficult, for example in OpenSSL AES128, only 0.00028\% of the encryption time relates to a target time \cite{Trautmann_Beckers_Wouters_Wildermann_Verbauwhede_Teich_2021} but several recent approaches use deep learning to help address this issue. Namely in \cite{Lu_Zhang_Cao_Gu_Lu_2021}, the authors propose a method to automatically focus the attention of the network on portions of traces that are too long to be used directly by existing methods.  Adversaries then analyze these to recover normally inaccessible information. Most traditional SCA on power and EM use Differential or Correlational Analysis, or a Template Attack.

\subsubsection{Differential Analysis}

In Differential Analysis, the goal is to create a selection function to classify traces into two sets \cite{DPA}. When a selection function properly distinguishes between criteria, there should be a notable difference between sets of traces. This selection function is a prediction of some intermediate state of the target algorithm. Typically, it uses a combination of known and unknown information. A possible value for the unknown information, such as the key, is guessed. The selection function uses this key guess to split traces into one of two groups. For each group, the average for each sample point is calculated from the members of the group. The difference between the resulting vectors (series of averages) is then found. A larger difference represents a more accurate selection function, while smaller differences indicate a poor one \cite{DPA}. An accurate selection function would have to be made with the correct unknown information, and thus confidential information can be recovered from traces. DPA remains an effective approach to SCA even against modern novel cryptographic algorithms \cite{Sim_Jap_Bhasin_2020}.

\subsubsection{Correlational Analysis}

Correlational Analysis also use the behavior of an intermediate state under hypothetical {key values}, but analyze the correlation of collected traces to this theoretical model rather than the differences between traces. An adversary creates a model to represent what a state would be for every possible value of the target information. For a cryptographic algorithm, the adversary would know the plaintexts, and using those, calculates intermediate state values for each value of the secret key. A matrix of the Hamming Weight (HW) or Hamming Distance (HD) of each state would then comprise the final model. HW and HD calculations are chosen as they represent power consumption \cite{kelsey2000hamming}. The more high voltage level bits (or changes in the the number of {high level} bits for HD) in a design, the more power is consumed. 
Next, one version of the model is compared against the measurement for a single time point across all traces. The Pearson Correlation between these two vectors is then calculated and recorded. This is then repeated for each time point and potential key pair. 
Finally, the highest correlation is found and the potential key that was used to create it is taken as the key guess \cite{CPA}.

\subsubsection{Template and Profiling Attacks}

Template attacks represent a fundamentally different approach to SCA from Differential and Correlational Analysis \cite{chari2002template}. In a profiling attack, a type of Template attack, an adversary has full control over a \textit{profiling device} which is identical to the victim device. The attacker then builds a model representing intermediate values using the profiling device, and uses that model to recover confidential information from the victim device. The structure of this model depends on a number of factors.  The adversary analyzes traces collected from the profiling device to identify components of the traces that distinguish the intermediate values. Traces from the victim device are then examined for these components and intermediate values predicted. For SCA on {software implementations of }AES, it is often a prediction of the state of a certain byte following the first substitution box operation. These predictions are then put through an inverse SBox and logical exclusive-or (XOR) operations with the plaintext to recover a key. Note that different algorithms may have their templates and models built and attacked in different manners. Templates can be built using either conventional \cite{Cabrera_Aldaya_Brumley_2021} or DL approaches \cite{yucross}.

\subsection{Side Channel Mitigation}

Due to how common side channels are and the threat they present, a number of defence techniques have been developed to protect against them. These can be divided into two main categories, Hiding and Masking.

\subsubsection{Hiding}

Hiding defences directly mitigate side channel leakage. This can be done through attenuating the signal and adding additional background noise~\cite{das2019stellar, das2018asni} or by randomizing time operations occur \cite{veyrat2012shuffling, boey2010random}. While this cannot entirely prevent leakage, these defences often increase the number of measurements required to perform a successful attack, thus restricting attackers. These approaches often present a trade-off between design complexity and side channel effectiveness. An example of this can be found in ``STELLAR" \cite{das2019stellar} where a method to almost completely mitigate electromagnetic (EM) side channel leakage is presented. However, the defence requires extremely accurate modeling of the EM behavior of a device, which is computationally complex. Circuitry to cancel out the EM emanations must then be designed, and placed above the protected design. On top of the complexity of this task, the protected design is restricted from using the top layers of the full design. These requirements greatly impede the design complexity and delay the time-to-market (TTM), although they do present significant security gains \cite{das2019stellar}. 

Another type of hiding technique is Dual-Rail Logic \cite{DRL}. In this, every gate has a partner gate with an inverse state (i.e. if the original gate is charged, the partner gate is discharged). This obviously incurred a heavy area and power overhead, as the number of gates in a design is doubled and power draw is kept at a steady elevated level. Dual-rail logic is highly effective at keeping power consumption steady and thus prevents accurate analysis of that side channel. However, other side channels, such as EM, are less effected \cite{immler2017your}. 

{It is also possible to mitigate side channel leakage through a process known as timing disarrangement \cite{timing_dis}. In this approach, designers purposefully add small amounts of clock jitter to different components. This jitter is not enough to affect the logical performance of the device, but by having the various components switch logic levels at different times, the side channel signal as a whole becomes distorted. This defence can easily be implemented through CAD and EDA tools automatically, though they have some performance overheads}

\subsubsection{Masking}

The other approach to side channel mitigation is to prevent a device from ever handling confidential information directly. This approach, known as masking, concedes that an attacker will be able to get accurate measurements from a device, and so instead prevents those measurements from being useful.  Byte masking has each operation split into subsections, combined with some other information (often called a mask), operated on, unmasked and recombined to restore the original data. The idea is to prevent adversaries from truly understanding what each operation involves, which significantly complicates side channel attacks. As the operations no longer match the adversaries models, they can no longer be easily recovered. 
However, it is possible to defeat masking with more complicated analysis \cite{peeters2005improved} and types of Machine Learning \cite{gilmore2015neural} to recover the mask used, after which the attacks can proceed as normal. Many types of machine learning do not use predicted models, and as such are mostly unaffected by masking. In fact, the most popular publicly available SCA dataset, ASCAD, includes traces from a byte-masked AES implementation on microprocessors \cite{ASCAD}. More machine learning based side channel analysis methods will be discussed later as the main focus of this survey paper.

\subsection{Deep Learning}

\begin{figure}
    \centering
    \includegraphics[width=0.5\columnwidth]{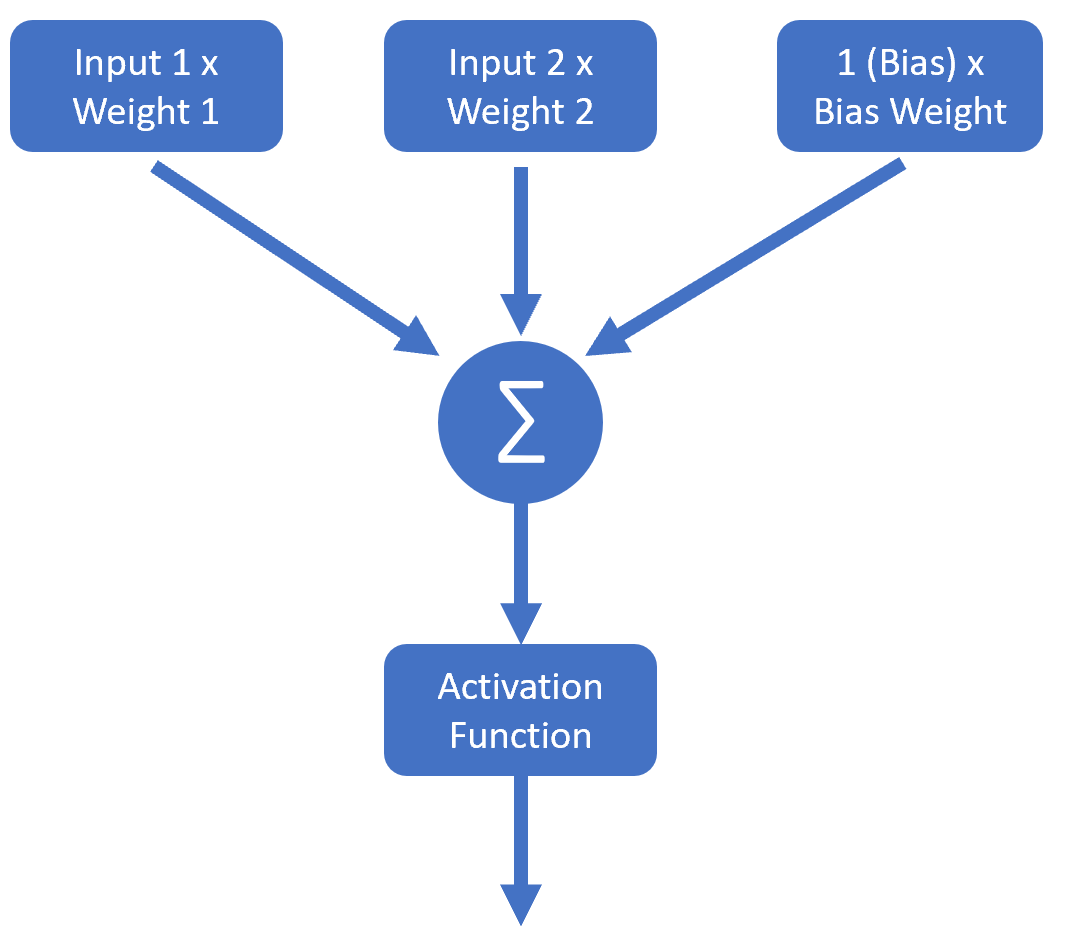}
    \caption{An example neuron in an artificial neural network.}
    \label{fig:neruon}
\end{figure}

\begin{figure}
    \centering
    \includegraphics[width=0.6\columnwidth]{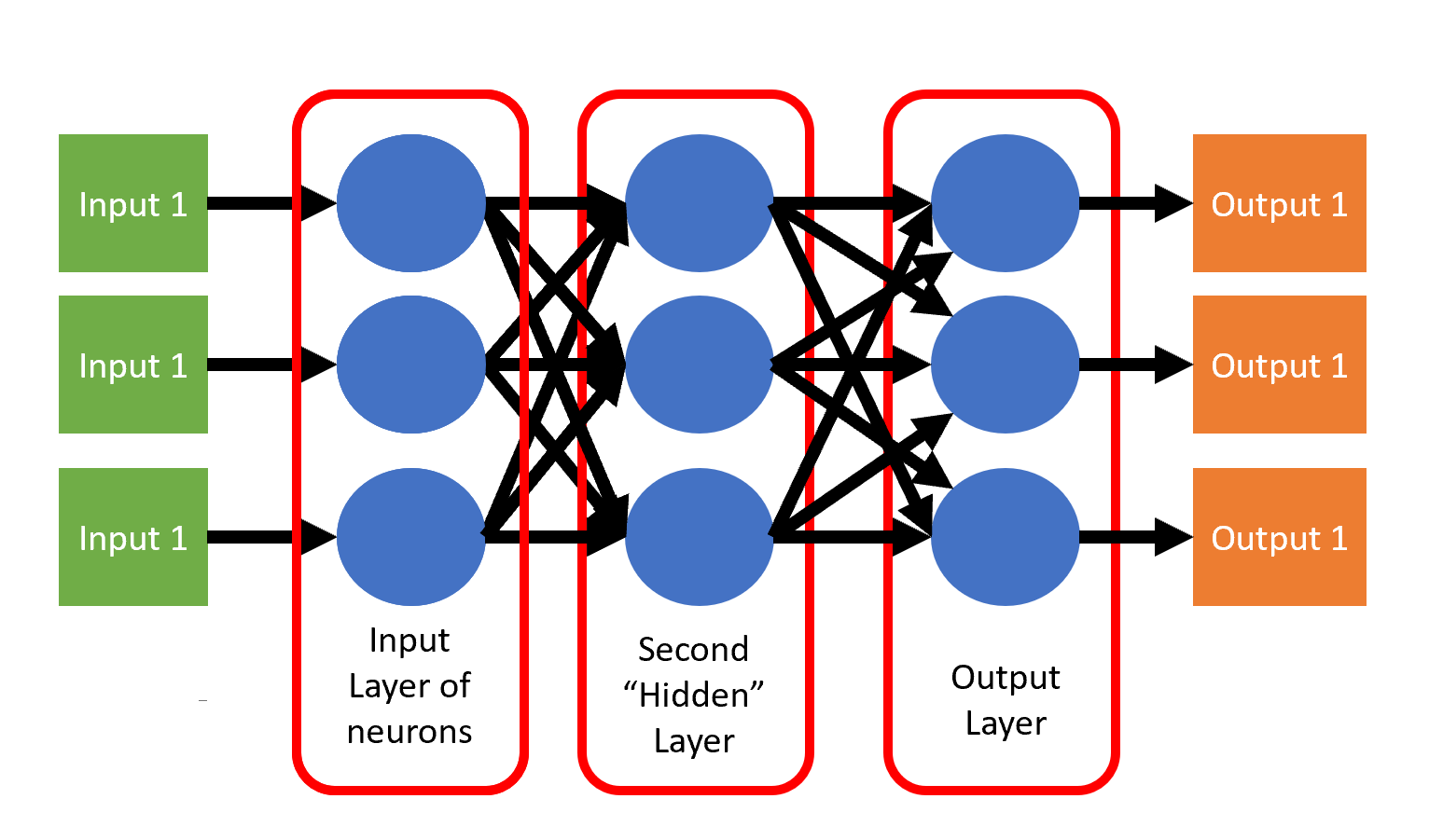}
    \caption{An example of a neural network with a hidden layer i.e. a deep learning network.}
    \label{fig:NN}
\end{figure}

Recent advances in machine learning, especially in deep learning, have made these techniques popular in other applications. 
Deep neural networks (DNN) are leading examples among all deep learning techniques.
DNNs consist of many neurons that ``activate'' under certain stimuli. In practice, this is often achieved by passing the weighted sum of inputs to an activation function that scales the result{, as shown in} Figure \ref{fig:neruon}. Deep Learning is so named as it includes artificial neural network using at least one ``hidden layer'' between input and outputs as seen in Figure \ref{fig:NN}.
Supervised learning is a type of deep learning where true outputs are known for all training inputs. Specifically,  The network takes in the input, calculates an output, and the output is then compared to the true output or \textit{label}. A \textit{loss function} quantifies the difference and the network adjusts its internal {parameters} based on the calculated loss, often through \textit{backpropagation}. This process is then repeated until the loss has been minimized and/or meet the requirements.

\subsubsection{Regression vs Classification Tasks}

Machine learning can either perform regression or classification tasks. In a regression task, the machine learning algorithm can take in outputs and extrapolate a numerical response to those inputs. An example of this may be an algorithm that takes in a car's mileage, age, model, and manufacturer and predicts the price it may sell for. 
Classification tasks on the other hand attempt to place a given input into one of several predetermined classes. Determining if a picture contains a cat or a dog would be a classification task. 

SCA with DL often falls into this category. There are a certain set range of possible byte values, which must be determined independently. 
For these values, known as ``labels'', the numeric distance between predicted and true labels is not indicative of correctness. This is to say if the true value for a certain key byte is 128, predictions of 129 and 34 are equally wrong. Thus, SCA for DL is often treated as a classification task.

\subsubsection{Multi-layer Perceptrons}

Multi-layer Perceptrons are the simplest form of DL. In an MLP architecture, many neurons or \textit{perceptrons} are layered and connected. An example of an MLP can be found in Figure \ref{fig:neruon}. Perceptrons that share inputs are called a \textit{layer}. There are at least two layers in a Deep Learning MLP, so that one layer is \textit{hidden}, or not exposed through the inputs or outputs of the model. It is possible for an MLP to {approximately} recreate any given function with enough perceptrons per layer \cite{csaji2001approximation}. However, instead of making wider MLPs (i.e. with more neurons per layer) to model complex functions, a more common approach is to make deeper MLPs (i.e. with more layers). It is important to note that in an MLP, layers can have different activation functions. 

For classification tasks, the final layer often has a softmax function as shown in Equation \ref{eqn:softmax}. $\vec{Z}$ denotes all the inputs to the neuron, which has $K$ outputs. $Z_i$ is the value of the $i$th element of $\vec{Z}$. The denominator scales the final outputs such that sum of them becomes one and thus each output represents a probability of that class. 

\begin{equation}
    \sigma(\vec{Z})_i = \frac{e^{Z_i}}{\sum^K_{j=1}e^{Z_j}}
    \label{eqn:softmax}
\end{equation}
\quad Rectified Linear Units (ReLUs) are a popular choice for layers other than the final {layer}. They prevent negative {activations} but allow positive ones, and have been found to be effective in a number of applications. Equation \ref{eqn:relu} demonstrates a ReLU. Y is the output and X is the input.
\begin{equation}
    Y = max(0, X)
    \label{eqn:relu}
\end{equation}

\subsubsection{Convolutional Neural Networks}
\begin{figure}
    \centering
    \includegraphics[width=1\columnwidth]{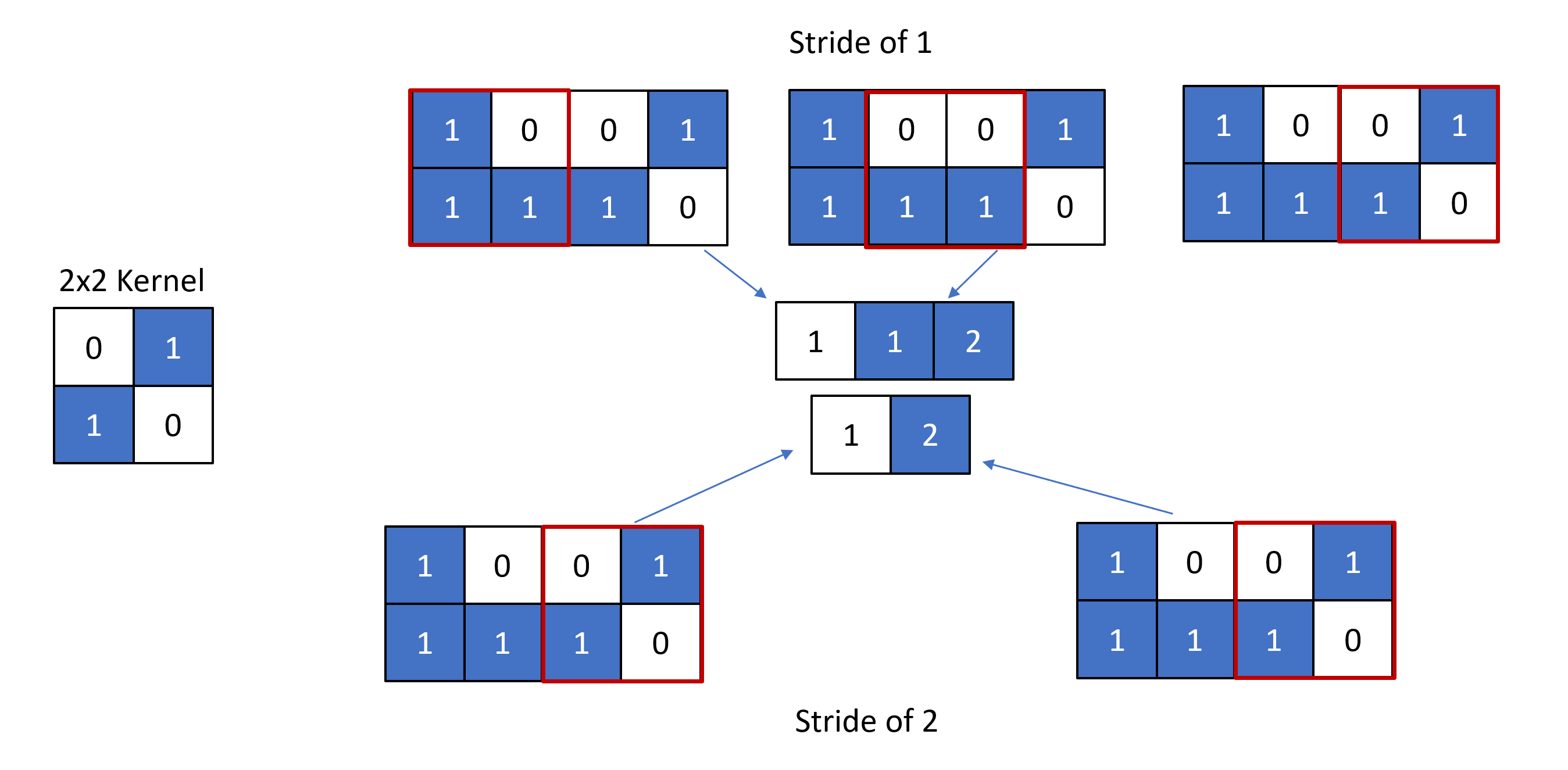}
    \caption{A simple example of kernel convolution.}
    \label{fig:conv}
\end{figure}

and
Convolutional Neural Networks (CNNs) are {often a} combination of convolutional layers followed by an MLP {when applied to classifications tasks such as in SCA. While others exist, such as Auto-Encoders, they tend to be auxiliary techniques}. Inputs to a CNN are first analyzed by convolutional layers, which attempt to identify patterns in the input data. The layer performs a convolution of a certain matrix, or \textit{kernel}, and the input to the layer. The kernel has a predetermined shape, and it moves across the input with a certain number of elements per step, or \textit{stride}. An element wise multiplication of the kernel and the {overlapping} portion of the input is performed, the results summed, and optionally fed through an activation function such as a ReLU. An example of this operation can be found in Figure \ref{fig:conv}. Convolutional layers can have multiple kernels, and are often used in combination to find different sized patterns, or patterns of patterns, which is a popular approach to computer vision \cite{lin2017focal}. This can be done by passing the outputs of one layer as the inputs to another, though without additionally precautions, this can quickly deplete the amount of information recovered in each layer \cite{szegedy2017inception}. The results of these convolutions are then passed into an MLP which results in the final output. Each kernel learns the weights/values of the elements that minimize the loss function, similar to a MLP.

\subsubsection{Auxiliary Layers}

In addition to perceptron (also known as fully connected) and convolutional layers, there are a few other types of layers that are important to deep learning. These are dropout, pooling, and batch normalization layers. Each fulfills an important role in a DL model. We will start by discussing dropout layers. These layers help prevent over-reliance on certain inputs or features, by randomly setting different features to zero during training. This obviously impedes the speed at which a model can learn, but it also tends to improve how well a model generalizes. If the model only needs a portion of the available information to reach the correct conclusion, then it tends to perform better when seeing new data that differs slightly from the training data. Pooling layers are often an integral component of CNNs. These layers are similar to convolutional layers {in that they operate by passing a kernel along their input}, but perform a \textit{pooling} operation instead of a convolution. Pooling operations distill several data points into a single one, often by finding the maximum value or taking the average of the set.
Finally, batch normalization or ``batch norm'' layers are essential parts of how modern DL models are structured, or their \textit{architecture}. Batch norm layers simply re-scale and re-center their inputs with each batch or set of inputs, to minimize artifacts that may affect model performance. 
All three are often used in state-of-the-art deep learning models \cite{lin2017focal, szegedy2017inception}.

\subsubsection{DL Architectures and Hyper-parameters}

DL models are more than just the weights of their parameters. The number and contents of each layer, how quickly weights update, and the way weight to change are identified are additional settings and known as \textit{hyper-parameters}. The number and contents of layers are known collectively as an \textit{architecture}. As mentioned in Section \ref{sec:background}, DL enhanced SCA often uses one of two architectures, an MLP or a CNN. MLPs tend to require fewer training traces and are easier to train, but are far more sensitive to synchronization issues than CNNs. There also more published guidance on creating CNN architectures for SCA~\cite{zaid2020methodology}. 

Ideal hyper-parameters can also be found through automated optimization or meta-learning. Automated optimization searches for ideal hyper-parameters through statistical analysis. ``Sherpa'' is one such example of this~\cite{hertel2020sherpa}. In Sherpa, ideal aspects of the model, often hyperparmeters are found by building a statistical model of the system's performance.
This model can take on many forms, from relatively simple Bayesian analysis to Markov chains \cite{gpyopt2016}. Most model-focused meta-learning follows a similar approach, though the statistical models are replaced with machine learned ones. This is the source of the name \textit{meta-learning} as a model learns how to best learn. Meta-learning can identify optimal architectures \cite{mishra2017simple} in additional to starting parameters \cite{finn2017model}. Hyper-parameter tuning methods for DL SCA have been proposed earlier including standard techniques such as Bayesian Optimization \cite{maghrebi2016breaking} \cite{Wu2020ICY}. However, novel approaches exist as well, as seen in \cite{Rijsdijk_Wu_Perin_Picek_2021} where the authors use a specific type of Deep Learning, known as Reinforcement Learning, to identify ideal model hyper-parameters. In Reinforcement Learning, an agent acts, evaluates the result of its actions, and quantifies a reward. This differs from Supervised Learning in that the reward value may not be known prior to the generation of the output (i.e. there is no ground truth label) but is only calculated after the agent completes its action. The agent then attempts to identify actions that maximize the reward.

\section{Contemporary Deep Learning Enhanced SCA}
\label{sec:prof}

As mentioned in Section \ref{sec:background}, profiling SCA naturally integrates well with supervised Deep Learning \cite{maghrebi2016breaking}. Each measurement, or trace, often belongs to a distinct operation, which can be used as a label. There is rarely any interdependency between measurements (i.e. each encryption is a stand-alone operation). Finally, the traces as captured are often time series, which DL historically works well on \cite{lecun1995convolutional}. This is not to say that this is the only way to perform DL SCA, in fact we cover several other approaches in Section \ref{sec:emerging}, but the majority of existing work on this topic follows this approach.

\subsection{Threat Model}

\begin{figure}
    \centering
    \includegraphics[width=0.6\columnwidth]{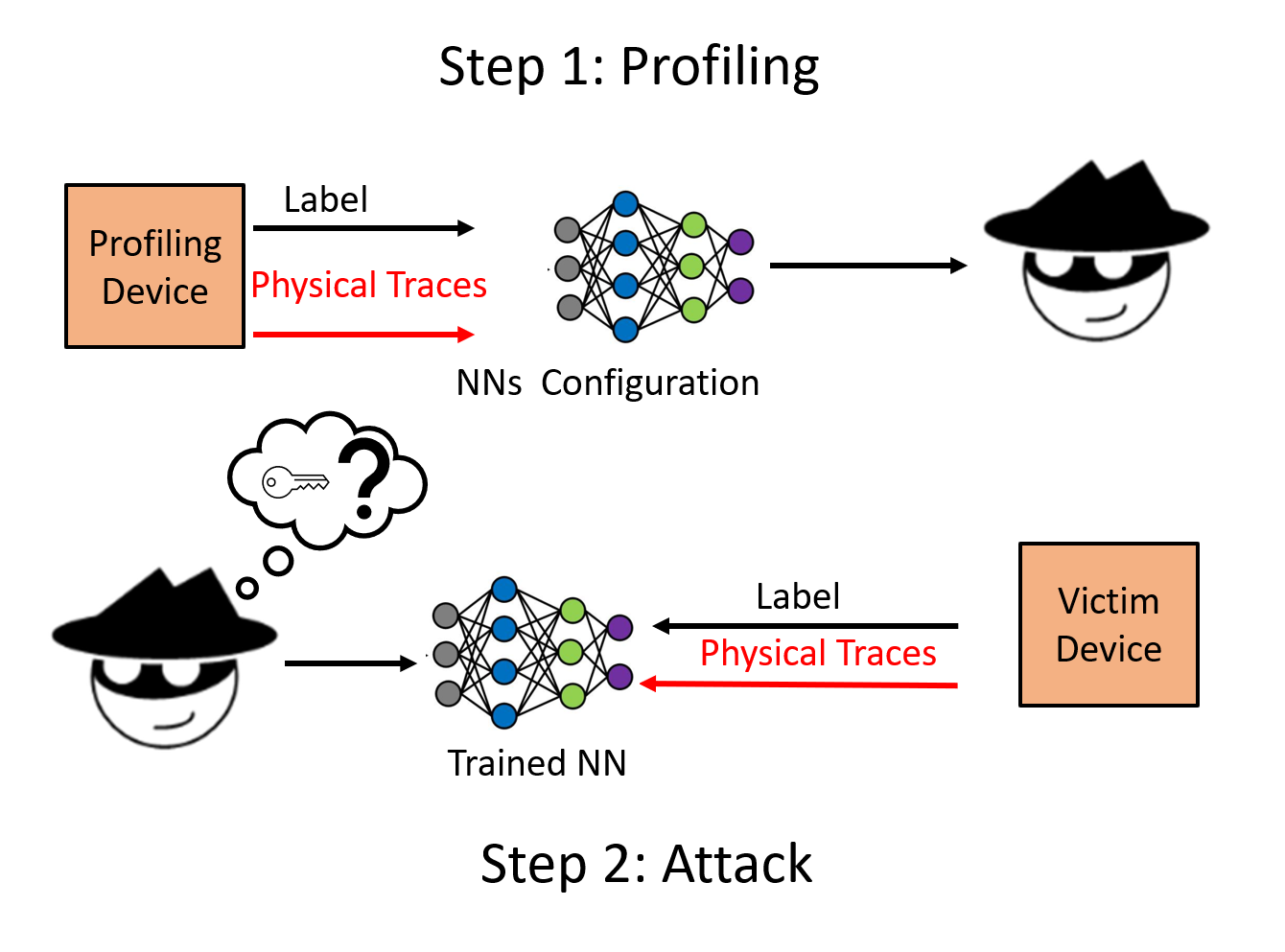}
    \caption{The two stages of a Profiling Attack. In Step 1, the adversary trains a neural network (NN) using a device with know internal states or labels. In step two, the attack uses the NN to recover the internal states or labels of an uncontrolled device.}
    \label{fig:prof_attack}
\end{figure}

In a profiled DL SCA attack, an adversary needs a target (i.e. confidential information to recover) and knowledge of a victim device (i.e. implementation or model information). We will take an adversary targeting the secret key used in AES as an example although DL SCA is not limited to AES \cite{Ngo_Dubrova_Guo_Johansson_2021}. In this case, an adversary knows that their victim device and knows the specific AES implementations, e.g., TinyAES running on an an ATMEGA microprocessor. 
Using this information, the adversary would then acquire an identical device, program it with identical code, and then collect traces from it under known inputs. The Substitution Box or \textit{SBox}, of the first round of AES is a common target. As the SBox operation is well defined and common to all AES implementations the adversary can easily reconstruct it. The inputs to the SBox of the first round is the result of XORing a byte of the secret key with the equivalent byte of plaintext (although it is possible to improve results by combining both plaintext and ciphertext \cite{Hoang_Hanley_ONeill_2020}). Thus, if the adversary has access to the plaintext and knowledge of the SBox output, they can recover the secret key. As such, the SBox output is often used as the label that a model attempts to learn during a profiled DL SCA attack. Said model would learn how to identify what SBox output a given trace contains. The adversary then needs to collect traces and plaintexts from the victim device, and uses the DL model to analyze those traces, thereby recovering the key. The general process can be found in Figure \ref{fig:prof_attack}

In more detail, two data sets must be collected from the device, one to use for profiling (the training dataset) and another to evaluate performance (the attack dataset). The attack dataset must have traces collected under the same key, but the profiling set can use random key-text combinations. The profiling data set is used to directly train the deep learning model as a \textit{training} set, though some portion of it may be reserved as a \textit{validation} set. This validation set is used to evaluate model performance while tuning parameters. The distinction between validation and test data is important to ensure unbiased analysis of the model's performance \cite{xu2018splitting}. {Once optimal parameters are found, the model is no longer updates it's internal parameters}, and then is supplied the attack or \textit{test} set for final analysis. A model's performance is evaluated thorough \textit{Guessing Entropy} (GE). GE can be calculated by following Algorithm 1 \cite{GE_orig}.

\begin{algorithm}
    \begin{algorithmic}[1]  
        \caption{Calculation of Guessing Entropy (GE) for a byte in the first round of AES. The set of traces to evaluate is $\mathcal{T}$. The set of plaintexts used to generate those traces is $\mathcal{P}$. These must be supplied in the same order as $\mathcal{T}$. The possible key values are $\mathcal{K}$. The number of evaluations to complete is $N_E$ while the number of traces per evaluation is $N_T$. $DL$ is a function representing the deep learning model. The true key, $K$, must also be supplied.}
            \REQUIRE { $\mathcal{T}, \mathcal{P}, \mathcal{K}$, $DL$, $K$, $N_E$, and $N_T$} \\
            \ENSURE  {Ranking of the correct key guess $R_K$}\\
            // First identify what key each SBox output relates to for a given plaintext
            \STATE $\mathcal{M}$ is a  $\lvert \mathcal{P} \rvert x \lvert \mathcal{K} \rvert$ matrix
           \FOR {$p$ in range($\lvert \mathcal{P} \rvert$)}
                \FOR {$k$ in range($\lvert \mathcal{K} \rvert$)}
                    \STATE $\mathcal{M}[p][k] = SBox(\mathcal{P}[p]~ xor~\mathcal{K}[k])$
                \ENDFOR
            \ENDFOR\\
            // Use $\mathcal{M}$ to find likelihood of $K$
            \STATE $R_E$ = matrix of size $N_E~x~N_T$
            \FOR {$e$ in range($N_E$)}
                \FOR {$t$ in range($N_T$)}
                    \STATE $i$ = random selection from range($\lvert \mathcal{T} \rvert$)
                    \FOR{$k$ in range($\lvert\mathcal{K}\rvert$)}
                        \STATE $\mathcal{K}[k]$ +=  $DL(\mathcal{T}[i])[k]$
                    \ENDFOR
                    \STATE $R_E[e][t]$ =  index of $K$ in reverse(sort($\mathcal{K}$))
                \ENDFOR
            \ENDFOR
            \FOR {$t$ in range($N_T$)}
                \STATE $R_K[t]$ = mean($R_E[all][t]$)
            \ENDFOR
    \label{alg:GE}   
    \end{algorithmic}
    
\end{algorithm}

\subsection{Intra-Device Profiled DL SCA}

In intra-device profiled DL SCA, the profiling and victim devices are the same physical device \cite{das2019x}. While there are legitimate threat models based around this methodology, intra-device attacks tend to be more useful as proofs-of-concept of a novel analysis algorithm rather than an implementable attack. This is not to say that such algorithms or the works using this are without merit. In fact, in many cases, only a few additional steps are required to adjust methodologies using intra-device data to work cross-device, which is the next topic we will discuss. 

Additionally, many of the publicly available databases for SCA use this approach. For example ASCAD, a popular public SCA database for AES implementations on microprocessors, consists of traces taken from a single microprocessor running a masked implementation of AES under varying conditions. This makes it an intra-device dataset. ASCAD has several subsidiary data sets, which allows it to account for a wide range of attack scenarios. Adversaries can build models for situations where a single fixed key is used during collection, or when both the key and plaintext change randomly. Additionally, while all traces in ASCAD are synchronized, the authors provide support for artificially de-synchronizing them, to simulate clock-jitter or poor testing conditions \cite{ASCAD}. These features have led to ASCAD being a popular publicly available database for SCA on microprocessors.

{A few examples of intra-device profiling being useful include Ranking Loss \cite{Zaid_Bossuet_Dassance_Habrard_Venelli_2020} and Multi-Leak Deep-Learning Side-Channel Analysis \cite{multi-leak}. These are two of the latest techiniques for DL SCA and they focus only on Intra-Device data. In Ranking Loss, the authors explore how changing the standard cross-entropy loss for one focusing on the rank of the true key in a Guessing Entropy attack (the true objective of a profiling attack) improve the model's performance. Multi-Leak takes a different approach, in that they look to combine the results of analyzing multiple leakage points simultaneously to recover the target key faster than other approaches which look at only a single source.}
\subsection{Cross-Device DL SCA}

While the above approach works well when analyzing the performance of a new type of analysis, methods using it often fail when the attacking and profiling devices are different instances \cite{das2019x, warriors}. This is likely due to slight differences in the side channel characteristics between the devices. This differences are best explained by the effects of process variation. As the distance between components affects resistance, capacitance, and other electrical characteristics, different physical instances of the same design may have slightly different side channel behaviors. 
There are two similar but distinct solutions to resolve the issue this causes for SCA.
Firstly, data can be collected from multiple profiling devices, and a DL model trained using the combination of all the data \cite{warriors, das2019x}. This has the benefit that models trained in this way are more likely to generalize to a new victim device, because the model already works for multiple distinct devices. However, there is no guarantee that it will work for the victim device, and a much large amount of training data is required. 

``X-DeepSCA" by Das et al. is one such approach \cite{das2019x}. In this work, the authors use multiple identical devices to create the profiling and attack datasets. They show that a DL model training on a single device fails to accurately predict a second device, but a similar model trained on traces from multiple devices (4 devices, 10k traces from each) can accurately (99.9\%) predict on an unseen device. In ``Mind the Portability'' Bhasin et al. take a similar approach, but also test different secret keys in addition to different devices \cite{warriors}.

The other approach is Transfer Learning (TL). In TL, a model is trained on a single source task normally. Then the model is \textit{fine-tuned} on the target task. Fine-tuning needs far fewer traces and training than training a model from scratch \cite{tranSCA}. In Cross Device TL for SCA, the adversary builds a model using a single profiling device. It is further assumed that the adversary has the access to the victim device and can obtain a limited number of traces (typically orders of magnitude fewer than needed to build a model) from it for known operations. The adversary then fine-tunes their model using these traces from the victim device. The resulting model can then quickly and accurately analyze new traces from the victim device \cite{tranSCA}. 

Thapar et al. demonstrate TL for SCA in ``TranSCA'' \cite{tranSCA}. They evaluate how many traces are needed to build a model from scratch as opposed to Transfer learning. They consistently find that far fewer traces are required for transfer learning than building a model \cite{tranSCA}. In fact, in one situation where a model needed 100,000 traces to perform well, less than one third of that was needed to transfer from another model. Obviously this requires a much more capable adversary than in in ``X-DeepSCA,'' but it also is significantly more efficient.

\section{Comparisons Among Deep Learning SCA}
\label{sec:Comparisons}

As many works use or at least evaluate their performance on the same database, ASCAD, we believe that is is fair to compare the performance reported between these methods. Through this comparison, we will evaluate how different approaches affect the performance of DL SCA. We will begin by comparing the Measurements to Disclose as reported in each work we have cited that obtained and reported conclusive results against synchronized traces from ASCAD dataset. These are the number of samples needed to recover the key, and as such lower numbers indicate a more successful or powerful attack method.

\begin{figure*}
    \centering
    \includegraphics[width=0.8\columnwidth]{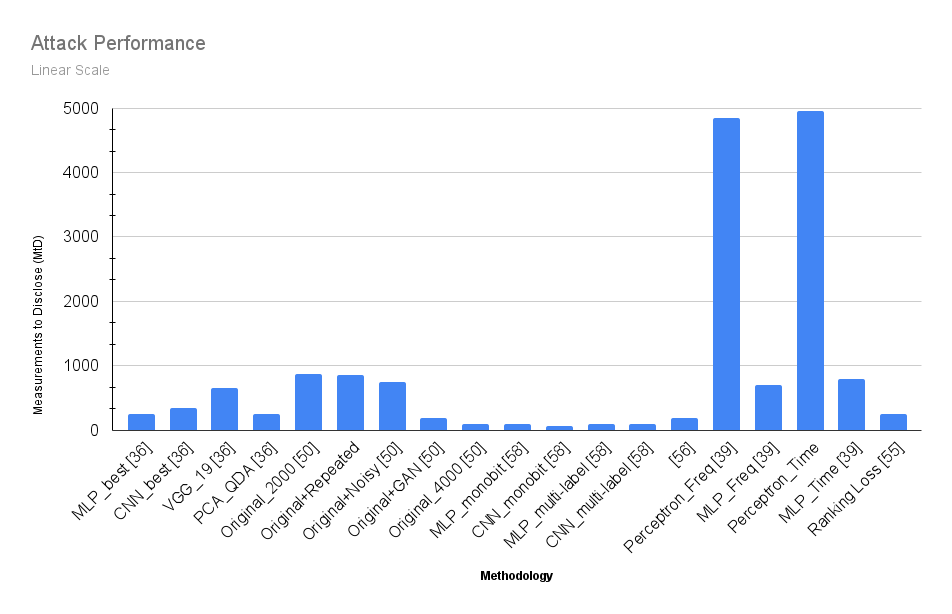}
    \includegraphics[width=0.8\columnwidth]{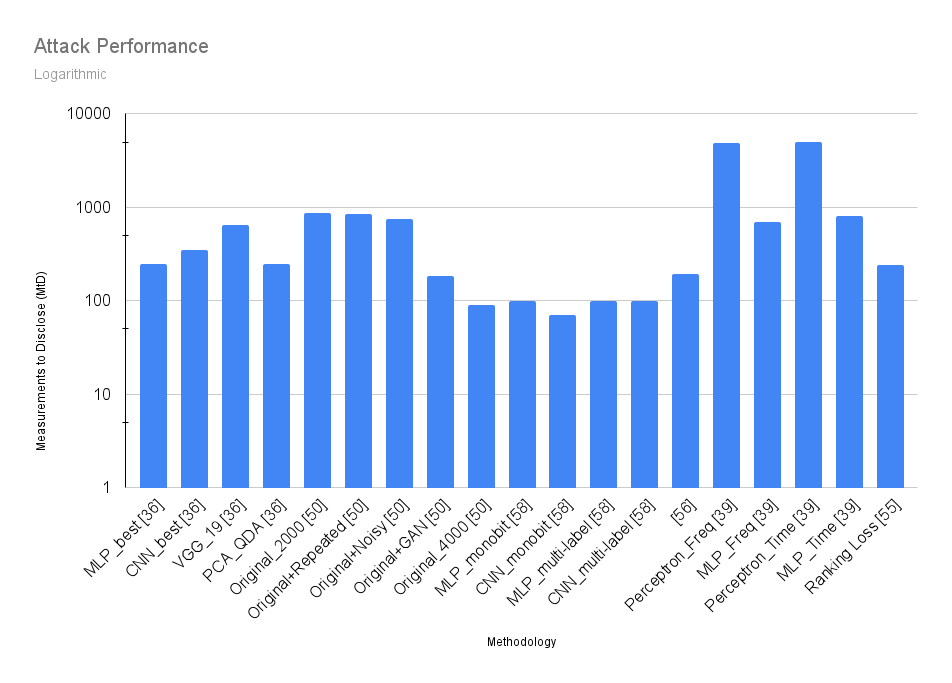}
    \caption{{Measurements to Disclose from the cited works which target ASCAD, lower is better. Each column is the name of the model provided by the work and its reference in our paper.}  The image above uses a linear scale, while the lower one uses a logarithmic scale, base 10.}
    \label{fig:GE_comp}
\end{figure*}

The results of this analysis can be found in Figure \ref{fig:GE_comp}. It should be readily apparent that some methods are more successful than others, with the single layer perceptron from \cite{multi-label} significantly under performing all other compared methods. Additionally we can see that the CNN monobit model from \cite{multi-label} has the best attack performance, needling only $\sim$70 attack traces to recover a key. We would also like to point out the PCA\_QDA method from \cite{ASCAD}, which is a statistical model (i.e. not a deep learning model) included as a baseline comparison.

Through the summary of how a created model functions, we believe that it would be useful to evaluate not just the performance of an attack, but also the difficulty of training a model. As such, we will perform an additional analysis using not only the Measurements to disclose (MTD) the correct as found through Guessing Entropy (GE, Algorithm \ref{alg:GE}), but a custom metric we call \textit{Key Recovery Difficulty} (\textit{KRD}). {Equation} \ref{eqn:RKD} showcases KRD, which will more fully represent the requirements of an adversary looking to deploy a given model, as both profiling and attack traces have a cost to collect. The cost of collecting these traces can are representing by $\alpha$ for the profiling traces and { the combination of }$\beta$ and $\gamma$ for the attacking traces. {This equation is derived through the observation that profiling and attack traces will be collected in different scenarios. When collecting profiling traces, the attacker is assumed to have complete control over the device, so profiling traces can be collected at a fixed cost ($\alpha$) per trace. When collecting attack traces however, the cost ($\beta$) is likely different and greater than when collecting profiling traces. Additionally, each additional trace collected presents a risk that the attacker may be discovered or additional time that the attacker needs to complete the attack. $\gamma$ can model how these costs grow with additional traces.}

\begin{equation}
    KRD = \alpha P_S + \beta A^{\gamma}
    \label{eqn:RKD}
\end{equation}

\begin{figure*}
    \centering
    \includegraphics[width=0.9\columnwidth]{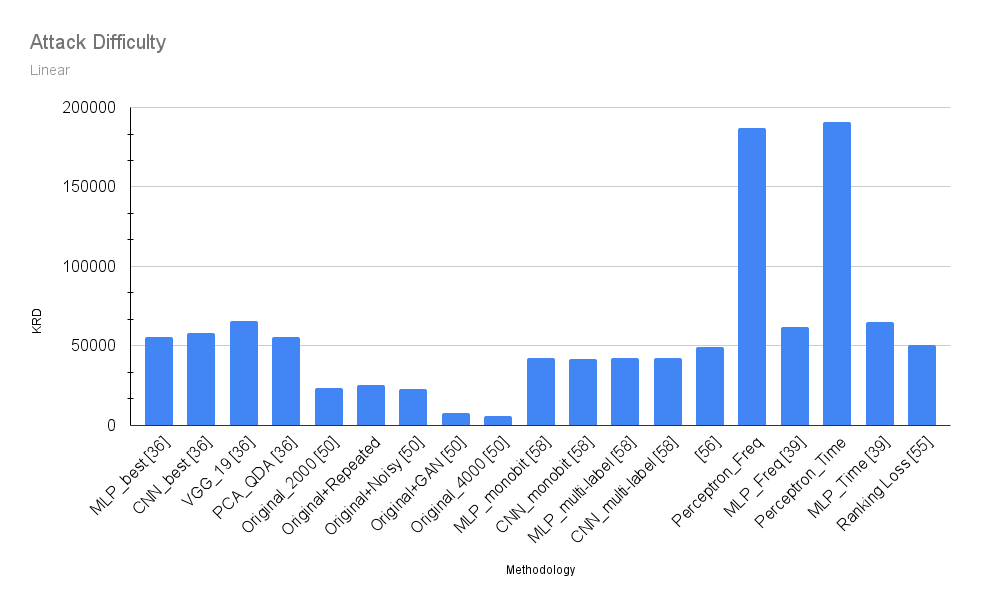}
    \caption{KRD for our cited works, lower is better.}
    \label{fig:RKD_comp}
\end{figure*}

$P_S$ represent the number of profiling traces used. $A$ is the number of attacking traces needed to recover the key through GE. If a method does not explicitly mention the number of traces from ASCAD used, we will assume that all of the appropriate set was (i.e. 60000 for profiling, 10000 for attack traces). 
{For the sake of demonstration,} we set $\beta$ to be 10 and $\alpha$ to be 1 in our analysis, which corresponds to attack traces being $10\times$ harder to collect than profiling. $\gamma$ is used to model the difficulty of collecting many traces from a victim device. We set this to be 1.1 in our analysis, the results of which can be found in Figure \ref{fig:RKD_comp}. It is important to note that in KRD, like GE, a lower score indicates a more powerful attack.

We can see that evaluating KRD instead of GE, the work from \cite{wang2020enhancing} is now the best. This is due to the authors of that works requiring far fewer real traces collected, as they artificially expand their data set. This is discussed more in Section \ref{sec:emerging}. As following that methodology requires far fewer real traces, it is easier for an attacker to accomplish, and potentially represents a higher threat.

\begin{table}[]
    \centering
    \caption{{Measurements to Disclose} from methods able to recover the key when traces had different levels of desynchronization {as defined in ASCAD. This is artificial jitter added to each trace, with 50 denoting a maximum of 50 sample windows of jitter and 100 denoting 100 sample windows}. \ding{55} means that the true key was never recovered.}
    \label{tab:desync}
    \begin{tabular}{|l|c|c|}
    \hline
        Method & De-sync 50 & De-sync 100 \\
        \hline
        CNN\_best \cite{ASCAD} & 4000 &  \ding{55}\\
        \hline
        PCA\_QDA \cite{ASCAD} & 4000 & \ding{55} \\
        \hline
        MLP\_monobit \cite{multi-label} & 5000 & \ding{55}\\
        \hline
        CNN\_monobit \cite{multi-label} & 200 & 500\\
        \hline
        CNN\_multi-label \cite{multi-label} & 150 & 350\\
        \hline
        \cite{zaid2020methodology} & 200 & 270\\
        \hline
        MLP\_Freq \cite{robyns2019improving} & 1000 & 1000\\
        \hline
        Ranking Loss \cite{Zaid_Bossuet_Dassance_Habrard_Venelli_2020} & 345 & 325 \\
    \hline

    \end{tabular}
\end{table}

However, there is one additional factor to consider.
All these results have been using synchronized traces, which may not represent real-world conditions well. In Table \ref{tab:desync} we show the effectiveness of all these techniques on ASCAD's synchronized traces. \ding{55} denotes that the method failed to recover the correct key.

\section{Emerging Directions for Deep Learning SCA}
\label{sec:emerging}

Profiling DL SCA has been well explored, leading to a wealth of information and understanding. However, there are several new areas of on-going research into DL SCA. These present a more robust understanding of the the strategies adversaries may deploy to increase their chances of success. As such, awareness and comprehension of these emerging areas is critical to any state-of-the-art analysis of DL SCA.

\subsection{Non-profiled Side Channel Analysis}

The first of these emerging areas concerns non-profiled DL SCA, or DL SCA that does not make use of profiling or template attacks. This has historically proven difficult for deep learning, as having accesses to well labeled data is a requirement to build a well performing model.
In \cite{timon2019non} Timon et al. exploit that requirement in order to recover a key from unlabeled traces. They propose Deep Learning Power Analysis or DLPA, which replaces the difference-of-means in Differential Power Analysis (DPA) with a deep learning algorithm. The central idea is that the adversary tries every single possible key value for a block in block cipher. 
Labels representing an intermediate value are then generated using the key value and and the model is trained using those labels. The best performing model is then assumed to have been trained using the correct key, as it was able to interpret some relationship between the data and labels. There are some obvious concerns with the current implementation, such as the dangers the model memorizing the data set, or poor architecture selection impacting the results, but nevertheless \cite{timon2019non} is a novel and insightful approach that turns one of DL's weaknesses into a useful aspect. This has been followed up on in recent years, most note ably with \cite{DDLA}, wherein the authors improve upon \cite{timon2019non} by introducing Deep Learning Leakage Assessment (DLLA). In this method, the statistical significance of side channel leakage is quantified through a deep learning approach rather than traditional methods.

Another approach to resolving DL SCA without profiling is proposed in ``Side-Channel Analysis with Unsupervised Learning'' (SCAUL) \cite{SCAUL}. Similar to the non-profiled analysis mentioned above, in SCAUL, the authors again modify a DPA attack for use with machine learning. However, they use a deep learning method to filter traces and create the intermediate state guesses, rather than for the final evaluation. Specifically, SCAUL uses an auto-encoder to down sample traces and select the most impactful points.
To do this, the auto-encoder takes in a trace, reduces it down to a smaller set of points, and then attempts to reconstruct the trace back to the original structure from those intermediate points. This selects the most important portions of a trace to use in DPA, rather than the entire set. This is an example of dimensional reduction. Those selected features are then classified by a MLP based on their predicted power draw and the results compared. If the secret key guess was correct, the difference between these two sets is high, while in all other scenarios, it is lower. As SCAUL does not require a known labels, it can be used in a non-profiled approach.

{One under explored area of non-profiled SCA is semi-supervised learning. Semi-supervised learning is a combination of unsupervised and supervised learning in which a small labeled dataset is used to enhance a larger unlabeled one. This could conceivably lead to a middle ground where a technique has most of the performance of a profiling attack, but can be conducted by a significantly less capable attacker.}

\subsection{Artificial Traces}

Another application of Deep Learning for Side Channel Analysis is to leverage DL to expand and improve datasets, rather than enhance the analysis process. This is often important to profiling and template attacks, both in deep learning and statistical implementations. These techniques require robust examples to model the behavior of a device under realistic sampling conditions. Collecting enough traces, and at a high enough resolution, to consistently accomplish this can be difficult. 
In ``Make some Noise'' \cite{kim2019makenoise}, Kim et al. suggest training a model on the ASCAD dataset, but adding artificial noise to each trace. They then compare how the amount of noise and number of affects the overall performance of the model. They find that additional noises improve model performance, especially when the test data set is already large \cite{kim2019makenoise}. 
This may be due to how the noisy traces belong to the same population as unseen but possible traces. Thus, these noisy traces are able to better represent the side channel behavior of the target device.

However, it is still unclear exactly how useful these artificial traces are. They then train a model on a combination of real and artificial traces and compare the performance to one trained on an equal number of real traces. This question is well addressed by Wang et al. in~\cite{wang2020enhancing}. In this work, the authors test how different ratios of artificially generated noisy traces and real traces impact the speed at which a secret key can be recovered. 
The authors also go even further and use a Generative Adversarial Network (GAN) to create new traces, instead of simply adding noise. Traces created through the GAN are found to have little to no significant difference on model performance when compared against real traces.

\subsection{Transferable Models}

Cross device attacks enable adversaries to train a deep learning model on a device they control and deploy it against one they do not. However, even in these situations, adversaries must know the device they are targeting and be able to acquire identical instances.
Through leveraging Model Agnostic Meta Learning (MAML), Yu et al. have developed a Meta-Transfer Learning technique that allows not only for cross-device but also cross-architecture attacks \cite{yucross}. In this threat model, an adversary has access to various devices, none of which are the same as their target device. Using these devices, the adversary collects a set of traces from each device. 
They then follow the steps described in Algorithm 2, treating each device's dataset as a task. 
First, a model is initialized with random weights, or \textit{parameters}. The state of the model is saved for later reference. For each task, a new model is initialized with those saved values, and the model is then trained on the task. The relative changes to parameters as needed to optimize that model are then applied to both the copy and the original model. This process is then repeated as many times as desired, so that the loss on the original model can be described as shown in Equation \ref{eqn:pre_training}.  Ultimately, this results in a model that requires the fewest changes to learn a new, yet unseen Task \cite{finn2017model, yucross}.  In \cite{Cao_Zhang_Lu_Gu_2021} Cao et al. propose a similar methodology using a custom loss based on Maximum Mean Discrepancy (MMD) instead of meta learning. In \cite{Zaid_Bossuet_Habrard_Venelli_2021}, Zaid et al. propose an approach combining multiple distinct trained models on Ensemble Loss. Specifically, this approach quickly combines separate machine learning models through a measure of their diversity.

There are still significant questions that remain in this field. Namely, those dealing with the limitations and optimal methodologies. The above works only show that transferable approaches are possible, and focus on microprocessors following similar (ARM cortex MX) architectures executing near identical C-code. It remains to be seen how different C-code, especially with masking, or different hardware, especially FPGA's or ASICs may affect these findings. In \cite {Cao_Zhang_Lu_Gu_2021} Cao et al. examine cross device FPGA SCA, using traditional 'mean discrepancy' metrics as a potential new direction. An evaluation of how transferable artificial traces may be, as well as whether the possibilities of `Meta-traces', or artificial traces designed for transferability have yet to be done as well.

\begin{equation}
\mathcal{L}^{D_i}_{\mathcal{T}_{i}}\left(f_{\theta}\right)=\sum_{\mathbf{x}(j), \mathbf{y}(j) \sim \mathcal{T}_{i}}l\left(f_{\theta}(x), y\right)
\label{eqn:pre_training}
\end{equation}
\begin{algorithm}
    \begin{algorithmic}[1]  
        \caption{MAML using tasks $\mathcal{T}={\{\mathcal{T}_{1}, \mathcal{T}_{2},...,\mathcal{T}_i}\}$, the DNNs classifier $f$ with parameter $\theta$, learning rate $\alpha$ and $\beta$ \cite{finn2017model}}
        \label{euclid} 
            \REQUIRE { $\mathcal{T}={\{\mathcal{T}_{1}, \mathcal{T}_{2},...,\mathcal{T}_i}\}$, $\alpha$ and $\beta$}
            \ENSURE  {Model parameters $\theta_M$}\\
            // DNNs model pre-training on the source task
            \STATE Randomly initialize parameter $\theta$
            \STATE Generate subset $D_t$ from the labeled dataset $D$ %
            \FOR {all $D_i$}
                \STATE Sample $D_i=(traces, labels)$ batch from dataset $D_t$
                \STATE Update parameter $\theta$ in Eq. (\ref{eqn:pre_training}) using Standard Gradient Decent 
            \ENDFOR \\
            // Meta-transfer learning on the target task
            \WHILE{not done}
                \STATE Sample task batch $\mathcal{T}_{i} \sim p(\mathcal{T})$ 
                \FOR {all batch $p(\mathcal{T})$}
                    \STATE Evaluate loss $\nabla_{\theta} \mathcal{L}_{\mathcal{T}_{\mathrm{i}}(\theta)}$ on sampled tasks
                    \STATE Optimize parameter $\theta_{\mathcal{T}_{i}}^{\prime} \leftarrow \theta-\alpha \nabla_{\theta} \mathcal{L}_{\mathcal{T}_{i}}(\theta)$ 
                \ENDFOR
                \STATE Update parameter $\theta \leftarrow \theta-\beta \nabla_{\theta} \sum_{\mathcal{T}_{\mathrm{i}} \sim p(\mathcal{T})} \mathcal{L}_{\mathcal{T}}\left(\theta_{\mathcal{T}_{\mathrm{i}}}^{\prime}\right)$
            \ENDWHILE
            \label{alg:MAML}
    \end{algorithmic}
\end{algorithm}

\subsection{Learning Embeddings}

AES is a common target of side channel Analysis targets AES due to both it being commonly used and because it is particularly vulnerable to ``divide-and-conquer'' strategies. {As AES operates on each of it's bytes independently.} adversaries can target smaller individual blocks rather than the whole state \cite{CPA}. These target blocks are often further broken down into individual bytes due to the nature of the AES algorithm and its Substitution Box in particular. These individual bytes can be brute forced with far greater easy than the entire state. A 128-bit AES implementation has 16 bytes each with 256 possible values, for a  search space with cardinatlity 4096. This is obviously smaller than the entire state space of $2^{128}$ values.

Cryptographers have made note of this flaw, and now design algorithms with larger blocks, often with 32 bits or more. Deep learning struggles to perform well on classification tasks with such a large range of potential values. Deep learning implementations often target individual bytes, but even that reduced space of 256 potential classes can impede performance \cite{picek2019curse}. In other fields, such as facial recognition where there may be millions of faces/classes in a dataset, deep learning uses embeddings instead of direct classes \cite{robyns2019improving}. {These embeddings are continuous representation of a large number of categorical values. The embeddings} are then analyzed to select a final recommendation. 

In \cite{robyns2019improving}, Robyns et al. suggest and implement an embedding based analysis for deep learning. They combine with with a custom loss function. This loss function actually maximizes the correlation between the results of the model's output with a traditional CEMA power model.
Through this method, the model does not predict all intermediate values directly, but rather hamming weight or hamming distance of those values. This significantly lowers the number of classes the model has to predict, and may enable deep learning classification on the latest cryptographic algorithms mentioned above. 

Additionally, they may greatly expand the scope of side channel attacks. Through embeddings, it may be possible to target much larger combinations of internal states than currently achievable. This obviously has implications to cryptographic algorithms which depend on having a large number of potential internal states for security, but also a number of different fields. For example, in side channel disassembly, attacks seek to revcover opcodes and operands being run a device through SCA. These each can have a substantial number of potential values, and even more when taken in conjunction (e.g. opcode and operand pairs). Embeddings present a possible solution to this increased complexity.

\section{Conclusion}
\label{sec:Conclusion}

Deep learning has numerous potential applications in hardware security, and especially in side channel analysis. In particular, profiling attacks on the power and Electromagnetic side channels using deep learning pose an increasing threat. While current defences may not be as effective against deep learning analysis as they are against traditional statistical analysis, deep learning has some inherent limitations. This is best demonstrated in the performance gap between intra- and cross-device attacks, though that gap is shrinking. Additionally, we provide a comparative analysis between different deep learning for side channel analysis methodologies, using the ANSSI SCA Database (ASCAD). The attack performance is evaluated through guessing entropy, and we provide an additional new metric, Key Recovery Difficulty (KRD), to quantify the combined final performance and the building of a deep learning model.

\begin{acks}

This material is based on research sponsored by Intel Corp., National Institute of Standards and Technology (NIST) and Office of Naval Research (ONR) Young Investigator Program (YIP). The U.S. Government is authorized to reproduce and distribute reprints for Governmental purposes notwithstanding any copyright notation thereon.
The views and conclusions contained herein are those of the authors and should not be interpreted as necessarily representing the official policies or endorsements, either expressed or implied, of Intel Corp., National Institute of Standards and Technology (NIST), Office of Naval Research (ONR) or the U.S. Government.

\end{acks}

\bibliographystyle{ACM-Reference-Format}
\bibliography{bibs/other_ml, bibs/sca_ml}

\end{document}